\documentclass[a4paper,12pt]{article}
\usepackage[left=2.5cm,bottom=3cm,right=2.5cm,top=3cm]{geometry} 

\usepackage{overpic,youngtab}
\usepackage{subfigure}
\usepackage[latin,english]{babel}
\usepackage{amsmath}
\usepackage{amssymb}
\usepackage{bigints}
\usepackage{epstopdf}
\usepackage{graphics,psfrag,rotating}
\usepackage{mathabx}
\usepackage{graphicx}
\usepackage{dcolumn}
\usepackage{float}
\usepackage{pdflscape}
\usepackage{array}
\usepackage{booktabs}
\usepackage{amscd} 
\usepackage{mathtools}
\usepackage{fancybox}
\usepackage{fix-cm}
\usepackage[colorlinks=true,citecolor=blue,,linktocpage=true,linkcolor=blue,urlcolor=black]{hyperref}
\usepackage{tikz} 
\usetikzlibrary{matrix} 
\usetikzlibrary{positioning} 
\usetikzlibrary{arrows}
\usepackage{titlesec}
\usepackage{abstract}

\def\be{\begin{equation}}
\def\ee{\end{equation}}
\def\bea{\begin{eqnarray}}
\def\eea{\end{eqnarray}}
\def\pd{\partial}

\def\xp{x^\prime}

\def\a{\alpha}
\def\b{\beta}
\def\g{\gamma}
\def\d{\delta}
\def\m{\mu}
\def\n{\nu}
\def\t{\tau}

\def\l{\lambda}

\def\r{\rho}

\def\bR{\bar{R}}

\def\bR{\bar{R}}

\def\s{\sigma}
\def\e{\epsilon}
\def\bi{\begin{itemize}}
	\def\ei{\end{itemize}}
\def\bg{\bar{g}}

\begin{document}
	
	\vspace*{-1cm}
\phantom{hep-ph/***} 
{\flushleft
	{{FTUAM-21-xx}}
	\hfill{{ IFT-UAM/CSIC-21-96}}}
\vskip 1.5cm
\begin{center}
	{\LARGE\bfseries  Gravitons in a gravitational  plane  wave. }\\[3mm]
	\vskip .3cm
	
\end{center}

\vskip 0.5  cm
\begin{center}
	{\large Enrique \'Alvarez, Jes\'us Anero and Irene S\'anchez-Ruiz.}
	\\
	\vskip .7cm
	{
		Departamento de F\'isica Te\'orica and Instituto de F\'{\i}sica Te\'orica, 
		IFT-UAM/CSIC,\\
		Universidad Aut\'onoma de Madrid, Cantoblanco, 28049, Madrid, Spain\\
		\vskip .1cm

		\vskip .5cm
		
		\begin{minipage}[l]{.9\textwidth}
			\begin{center} 
				\textit{E-mail:} 
				\tt{enrique.alvarez@uam.es},
				\tt{jesusanero@gmail.com} and
				\tt{irenesanchezl2@gmail.com}
			\end{center}
		\end{minipage}
	}
\end{center}
\thispagestyle{empty}

\begin{abstract}
	\noindent
Gravitational plane waves (when Ricci flat) belong to the VSI family. The achronym VSI stands for  {\em vanishing scalar invariants}, meaning that all scalar invariants built out of Riemann tensor and its derivatives vanish, although the Riemann tensor itself does not. In the particular case of plane waves many interesting phenomena have been uncovered for strings propagating in this background. Here we comment on  gravitons propagating  in such a spacetime.
\end{abstract}

\newpage
\tableofcontents
\thispagestyle{empty}
\flushbottom

\newpage

\section{Introduction.}
The exact solutions of Einstein equations that can be in some sense interpreted as gravitational waves are interesting in their own right. What is more, the simplest of those, namely plane waves, is in some sense the limit of an arbitrary spacetime  in the vicinity of a null geodesic (this is called the {\em Penrose limit}) \cite{Penrose}.
\par
Into the plane waves, there is an   interesting subfamily of spacetimes \cite{Jordan} and \cite{Deser},  to wit, that of a special class of exact gravitational waves, namely  the {\em plane-fronted parallel waves with parallel rays} (PP waves) that belong to the {\em vanishing scalar invariants} (VSI) class of spacetimes.  That is, all scalars constructed out of the metric, the Riemann tensor and its derivatives vanish identically. This is obviously true also in flat spacetime where Riemann tensor vanishes. The interesting thing is that can be also true in {\em non flat spacetimes}, with a non vanishing Riemann tensor.

\par
This fact means that there are no allowed counterterms \cite{Gibbons} in the Einstein-Hilbert action when considering gravitons propagating in such a background.  Which has led some authors to claim that there are no divergences whatsoever, and quantum gravity must be  finite around such spaces. This is not so, and it is enough to remember that Minkowski  is just one such space. Quantum gravity in flat space (and its divergences) has been investigated by Capper and coworkers a long time ago \cite{Capper}.
\par
 The S-matrix elements (on shell divergences) however should be finite \cite{Kallosh}. There is a general statement which asserts  that divergences that vanish on the equations of motion  for the quantum fields (as they do in VSI) indicate that modifications of the BRST algebra are necessary  at each loop order, but we are not elaborating on that in this paper.

\par
Our aim here instead is to consider some aspects of the physics of such gravitons propagating in a plane gravitational wave.  
\par
We review some usuals definitions (see Appendix \ref{A} for more details). The standard {\it pp-wave} metric reads
\be
\text{d}s^2=\text{d}u\text{d}v-F[u,x^c]\text{d}u^2-\d_{ab}\text{d}x^a\text{d}x^b
\ee
where $u\equiv t-z$, $v\equiv t+z$,  and the transverse coordinates are $x^c=(x,y)$.
These spacetimes belong to the Kerr-Schild class \cite{Kerr}. This means that the metric can be written as
\be g_{\m\n}=\eta_{\m\n}+F l_\m l_\n\ee
with
\be l_\m=-\d_{u\m}\ee
and the additional facts that $l^2=0$, $R_{\m\n\r\s} l^\s=0$ and $\pounds(l) R_{\m\n\r\s}=0$. An interesting fact proved in \cite{Gurses} is that Einstein equations can be written in a linear form and the pseudoenergy tensor vanishes.

A {\it plane wave} is a classical pp-wave for which the characteristic function $F$ is quadratic in $x^c$, $F[u,x^c]=H_{ab}x^ax^b$ 
\be\label{Bm}
\text{d}s^2=\text{d}u\text{d}v-H_{ab}[u]x^ax^b\text{d}u^2-\d_{ab}\text{d}x^a\text{d}x^b
\ee
in harmonic (often called Brinkmann) coordinates.

\par
This metric has an  only nonvanishing component of the Riemann tensor 
\be
R_{uaub}=- H_{ab}[u]
\ee
and the Ricci tensor, in turn, has reduced to
\be
R_{uu}=-\d^{ab}H_{ab}[u]\equiv -H^a_a[u]
\ee
Obviously the manifold is Ricci flat whenever the transverse matrix $H_{ab}$ is traceless, and the matrix  is usually written as
\be
H_{pq}[u] x^p x^q \equiv a_+[u](x^2-y^2)+2 a_\times[u]  xy
\ee
here the functions $a_+[u]$ and $a_\times[u]$ are the {\em polarization profiles } of the wave.

In the non-Ricci-flat  case, there is a source 
\be
T_{uu}={1\over 4}\left(T_{tt}+T_{zz}\right)-{1\over 2} T_{tz}=-{1\over \kappa^2} H^a_a[u]
\ee


Finally, an especially interesting case  are the {\em impulsive waves} \cite{Garriga:1990dp} and \cite{Khan:1971vh}, which we will study in the next section
\be
\text{d}s^2=\text{d}u\text{d}v-\frac{1}{\l_a}\d(u)\,\d_{ab}x^ax^b\text{d}u^2-\d_{ab}\text{d}x^a\text{d}x^b
\ee
this metric is Ricci flat if
\be \l_1+\l_2=0\ee
\par
We shall see in a moment that in some cases, it is convenient to transform  to a set of {\em group coordinates} or Rosen coordinates \cite{Garriga:1990dp}, where the metric reads
\be
\text{d}s^2=\text{d}u\text{d}V-h_{ab}[u] \text{d}X^a\text{d}X^b
\ee

The relationship between the two sets of coordinates is given by 
\bea
&&v=V+\frac{1}{2}\dot{h}_{ab}(u)X^aX^b\nonumber\\
&&x^a=P^a_b(u)X^b
\eea
with
\be  h_{ab}(u)=\d_{ij} P_a^i(u)P_b^j(u)\ee
where the matrix $P_a^b(u)$ is determined by solving the differential equation
\be \d_{ij}\ddot{P}^i_a(u)=H_{kj}(u)P^k_a(u)\ee
with  initial conditions satisfying the constraint
\be \d_{ij}\left(\dot{P}^i_a(u)P^j_b(u)-\dot{P}^i_b(u)P^j_a(u)\right)=0\ee
where the overdot means derivative with respect to $u$.

These coordinates do not cover the whole manifold because the determinant of the transverse metric vanishes at some point $u=\overline{u}$.
The only nonvanishing component of Ricci tensor in this coordinates reads
\be
R_{uu}=-{1\over 2} h^{ab}\ddot{h}_{ab}-{1\over 4} \dot{h}^{ab}\dot{h}_{ab}
\ee
There is a theorem, first pointed out in \cite{Horowitz}, which states that imposing additional Ricci flatness, then all higher order corrections to the equation of motion also vanish. This means the exact equation of motion of the full perturbative effective action. \\

\par
The most important aspect of VSI spacetimes is that  all the DeWitt-Schwinger coefficients of the short time expansion of the heat kernel do necessarily vanish. We will expose this in some  detail in the correspondent section.\\
\par
Is there any physical meaning in the fact that such an action vanishes on shell? 
In fact there is none. For example, any theory characterized by a bilinear Lagrangian
\be
\mathcal{L}=\sum \phi_i M_{ij} \phi_j
\ee
however complicated the kernel $M_{ij}$ also vanishes on shell (like Dirac Lagrangian for example).
\par
 In addition, contrary to popular belief,
this is {\em not} a generic property  of covariant actions. A simple example, the Weyl squared action
\be S\equiv \int d^4x\sqrt{|g|}W_{\m\n\a\b}W^{\m\n\a\b}\ee
which equation of motion is the vanishing of Bach tensor, but they  do not imply $W_{\m\n\r\s}^2=0$. That is, the Lagrangian of conformal gravity does not vanish on shell. 
\par
The fact that it does for Einstein is just an accident.
Another point is that of course this does {\em not} mean that the equation of motion are trivial.

The preceding theorem shows that

\bea
&&\left.\Gamma[g]\right|_{on\,shell}=\Gamma_{class}[g]=-{1\over 2 \kappa^2} \int d^4x\sqrt{|g|}R\nonumber\\
&&\left.{\d \Gamma[g]\over \d g_{\a\b}(x)}\right|_{on\,shell}=0
\eea
We have just asserted that this last fact has been shown to be true in VSI backgrounds by Horowitz and Steif \cite{Horowitz}; it  physically means   that the VSI spacetime is the consistent vacuum of quantum gravity.
But of course  the 1PI two point correlator is different to zero.
\be
\left.{\d^2 \Gamma[g]\over \d g_{\a\b}(x)\d g_{\b\g}(\xp)}\right|_{on\,shell}\neq 0
\ee

In the next sections we will show our most relevant results about VSI spacetimes and we will do some illustrative examples.

\section{The heat kernel for a plane wave.}
The most useful tool in order to compute both the propagator as well as the effective action is the heat kernel \cite{AA}. In order to do this calculation it pays to work in Rosen coordinates
\be
ds^2= du dV-h_{ab}[u] dX^a dX^b
\ee 
We shall see in a moment that the propagator is the inverse of the plane wave d'Alembertian operator. In fact, the effective action  will be determined by the determinant of the same operator. To be specific the  d'Alembertian for VSI spacetime reads
\bea &&\Box=\frac{\dot{h}_{ab}}{h_{ab}}\partial_v+\partial_u\partial_v-h^{ab}\partial_a\partial_b
\eea

There is a theorem by DeWitt \cite{AA} on the existence of a small proper time expansion for the heat kernel of any operator which can be obtained by a  deformation of the d'Alembertian,   given by
\be
K\left(\tau;x,y\right)=K_0 \left(\tau;x,y\right)~\sum_{p=0}^\infty~a_p \left(x,y\right)\tau^p
\ee
with 
\be a_0(x,y)=1\ee
and the $K_0\left(\tau;x,y\right)$ is the flat space heat kernel
\be
K_0 \left(\tau;x,y\right)={1\over (4 \pi\t)^{n\over 2}}\,e^{-{\s(x-\xp)\over 2\t}}
\ee
Here $\s(x,\xp)$ is the world function corresponding to the points $x$ and $\xp$, \cite{Synge:1960ueh}.
\par
 The important thing for our purposes is that the so called DeWitt-Schwinger coefficients, $a_p$ are given  by integrals of  pointwise scalars constructed out of Riemann tensor and its covariant derivatives contracted with the metric tensor. But in all VSI metrics, including plane waves, there are none of those. We are forced to conclude that 
 \be
 a_p=0\hspace{0.5cm}\text{if} \, p\neq 0
 \ee
 and the heat kernel reduces to flat space one
\be
K\left(\tau;x,y\right)=K_0 \left(\tau;x,y\right)
\ee
this is a {\em exact} expression for the heat kernel. \\

We want to emphasize that we only need to obtain the world function to  know the {\em exact} heat kernel.\\

In order to compute  the world function, we consider the  Lagrangian 
\be \mathcal{L}=\frac{du}{d\t}\frac{dV}{d\t}-h_{ab}[u]\frac{d X^a}{d\t}\frac{d X^b}{d\t}\ee
where $\t$ is the proper time. The canonical momenta read
\bea 
&&p_u=\frac{dV}{d\t}\nonumber\\
&&p_V=\frac{du}{d\t}\nonumber\\
&&p_a=-2h_{ab}\frac{d X^b}{d\t}
\eea
The Euler-Lagrange equation corresponding to the coordinate  $V$ implies the conservation of $p_V$. We can then write
\be u-u_0=p_V(\t-\t_0)\ee
The transverse momentum can be written as
\be p_a=-2h_{ab}p_V\frac{dX^b}{du}\ee
therefore
\be X^a=X_0^a-\frac{p_b}{2p_V}\int^u_{u_0} h^{ab}[u']du'\ee
The first integral of the timelike geodesic equations
\be
g_{\m\n}{d x^\m\over d\t}{dx^\n\over d\t}=1
\ee
implies
\be \frac{du}{d\t}\frac{dV}{d\t}-h_{ab}\frac{dX^a}{d\t}\frac{dX^b}{d\t}=\frac{dV}{du}p^2_V-\frac{1}{4}h^{ab}p_ap_b=1\ee
so that

\bea\label{gt}
&&V=V_0+\frac{u-u_0}{p_V^2}+\frac{p_ap_b}{4p_V^2}\int^u_{u_0} h^{ab}[u']du'
\eea

Finally, the world function reads
\bea \s(x,x_0)&&=\frac{1}{2}(\t-\t_0)^2=\frac{1}{2p_V^2}(u-u_0)^2=\nonumber\\
&&=\frac{1}{2}(V-V_0)(u-u_0)-\frac{p_ap_b}{8p_V^2}(u-u_0)\int h^{ab}[u']du'=\nonumber\\
&&=\frac{1}{2}(V-V_0)(u-u_0)-\frac{1}{2}(X^a-X^a_0)C_{ab}[u,u_0](X^b-X^b_0)(u-u_0)\eea
where, we can define $C_{ab}$  as
\be\label{1} C_{ab}[u,u_0]\int_{u_0}^u h^{cb}[u']du'=\d_a^c\ee

In conclusion, with the characteristic function $h_{ab}[u]$ of the metric, we can obtain the world function and therefore the heat kernel. We show an explicit example





\subsection{Impulsive Gravitational Plane Waves.}
A particular case worth considering is the {\it impulse gravitational plane waves}
\be
\text{d}s^2=\text{d}u\text{d}v-\frac{1}{\l_a}\d(u)\d_{ab}x^ax^b\text{d}u^2-\d_{ab}\text{d}x^a\text{d}x^b
\ee
which in Rosen coordinates translates to
\be
\text{d}s^2=\text{d}u\text{d}V-\left(1-\frac{u\Theta[u]}{\l_a}\right)^2\d_{ab}\text{d}X^a\text{d}X^b
\ee
where $\Theta[u]$ is the Heaviside step function. This physically represents two 
flat spaces glued together at $u=0$ , one corresponding
to $u<0$ in Minkowskian coordinates and the other corresponding to $u>0$ in non-Minkowskian ones.

The world function, using our previous results, reads
\bea \s(x,x_0)=\frac{1}{2}(V-V_0)(u-u_0)-\frac{1}{2}(X^a-X^a_0)C_{ab}[u,u_0](X^b-X^b_0)(u-u_0)\eea
where
\be C_{a}^{b}[u,u_0]=\frac{1}{\int^u_{u_0} \left(1-\frac{u'\Theta[u']}{\l_a}\right)^2du'}\d_{a}^b\ee

To be specific, 
\begin{itemize}
	\item $u_0,u<0$
	\be C_{a}^{b}[u,u_0]=\frac{1}{(u-u_0)}\d_{a}^b\ee	
	\item $u_0<0$ and $u>0$, in this case
	\be \int^u_{u_0} \left(1-\frac{u'\Theta[u']}{\l_a}\right)^2du'=\int^0_{u_0} du'+\int^u_{0} \left(1-\frac{u}{\l_a}\right)^2du'=u-u_0 - \frac{u^2}{\l_a}+\frac{u^2}{3\l_a}\ee
	then
	\bea C_{a}^{b}[u,u_0]&&=\frac{1}{u-u_0 - \frac{u^2}{\l_a}+\frac{u^2}{3\l_a}}\d_{a}^b\eea
	\item $u_0,u>0$, now
	\be \int^u_{u_0} \left(1-\frac{u'}{\l_a}\right)^2du'=-\frac{\l_a}{3}\left(1-\frac{u}{\l_a}\right)^3+\frac{\l_a}{3}\left(1-\frac{u_0}{\l_a}\right)^3\ee
	then
	\be C_{a}^{b}[u,u_0]=\frac{1}{-\frac{\l_a}{3}\left(1-\frac{u}{\l_a}\right)^3+\frac{\l_a}{3}\left(1-\frac{u_0}{\l_a}\right)^3}\d_{a}^b\ee
\end{itemize}
this world function has a singularity when $u=\l$ (where the determinant $h=0$) which marks the boundary of the normal neighborhood. This is the place where the nearest zero of the Jacobi field is located.

\section{The propagator of the graviton field.}
Let us begin with a general remark.  In the reference \cite{Carmelo}
a  general parametrization  of the free graviton propagator in flat space is introduced, namely
\begin{equation}
\begin{array}{l}
{\langle h_{\m_1\m_2}(k) h_{\m_3\m_4}(-k)\rangle_0=}\\[4pt]
{\,i\dfrac{A_1}{k^2}\left(\eta_{\mu_1\mu_3}\eta_{\mu_2\mu_4}+\eta_{\mu_1\mu_4}\eta_{\mu_2\mu_3}-\eta_{\m_1\m_2}\eta_{\m_3\m_4}\right)
+\,i\dfrac{A_2}{k^2}\eta_{\mu_1\mu_2}\eta_{\mu_3\mu_4}
+\,i\dfrac{A_3}{(k^2)^2}\left(\eta_{\mu_3\mu_4}k_{\mu_1} k_{\mu_2}+\eta_{\mu_1\mu_2}k_{\mu_3} k_{\mu_4}\right)}\\[4pt]
{+\, i\dfrac{A_4}{(k^2)^2}\left(\eta_{\mu_1\mu_3}k_{\mu_2} k_{\mu_4}+\eta_{\mu_1\mu_4}k_{\mu_2} k_{\mu_3}+\eta_{\mu_2\mu_3}k_{\mu_1} k_{\mu_4}+\eta_{\mu_2\mu_4}k_{\mu_1} k_{\mu_3}\right)
+ \,i\dfrac{A_5}{(k^2)^3} k_{\mu_1} k_{\mu_2} k_{\mu_3} k_{\mu_4}.}
\end{array}
\label{genprop}
\end{equation}
$A_i$, $i=2..5$ are  constants. We shall assume that
\begin{equation}
A_1=\dfrac{1}{2}
\end{equation}
which is just a normalization.
The contribution
\begin{equation}
G_{\mu_1\mu_2\mu_3\mu_4}\equiv \,\dfrac{1}{p^2}\left(\eta_{\mu_1\mu_3}\eta_{\mu_2\mu_4}+\eta_{\mu_1\mu_4}\eta_{\mu_2\mu_3}-\eta_{\m_1\m_2}\eta_{\m_3\m_4}\right)\equiv {1\over p^2} P_{\mu_1\mu_2\mu_3\mu_4}
\end{equation}
to the propagator can be {interpreted} as coming from the bit of the graviton field which contains the physical graviton polarizations and the corresponding creation and annihilation operators. This is in fact exactly the propagator used in \cite{Capper}.
\par
This propagator has several interesting properties. First of all, there is a unique pole at $p^2=0$, and the residue
for $G_{00,00}$ is positive, although it is neither traceless not transverse.
\par
The need to recover the Newtonian potential in the static case impose some restrictions. In fact, 
using (\ref{genprop}), one readily deduces that
\begin{equation}
\left\langle\hat{h}^{00}(k)\hat{h}^{00}(-k)\right\rangle_0=-\dfrac{i}{4\vec{k}^2}\big[3+\dfrac{1}{4} A_5+ A_4\big],
\end{equation}
for static sources, namely,  $k^\m=(0,\vec{k})^\mu$. Inserting the previous result in the static  potential
\begin{equation}
V_{\rm Nw}(\vec{k})=-i\dfrac{1}{4}m_1 m_2 \left\langle\hat{h}^{00}(k)\hat{h}^{00}(-k)\right\rangle_0=-\dfrac{m_1 m_2}{16\,\vec{k}^2}\big[3+\dfrac{1}{4} A_5+ A_4\big]
\label{NWAMP}
\end{equation}

We learn that in order to recover the correct potential, 
\begin{equation}
V_{\rm Nw}(\vec{k})\,=\,-\dfrac{1}{8}\dfrac{m_1 m_2}{\vec{k}^2},
\label{NWPOT}
\end{equation}

we need that 
\begin{equation}
A_4+\dfrac{1}{4} A_5=-1,
\label{linearconstraint}
\end{equation}

We thus conclude that both $A_4$ and $A_5$ in the free graviton propagator in (\ref{genprop}) cannot vanish at the same time, { regardless} of the Lorentz covariant gauge-fixing term that one uses. \footnote{ Let us here report the corresponding propagator in unimodular gravity. Here quartic poles appear, although they do not couple to the traceless part of the energy-momentum tensor \cite{Trees}

\bea
&&G_{\mu_1\mu_2\mu_3\mu_4}(p)\equiv\langle h_{\mu_1\mu_2}(-p) h_{\mu_3\mu_4}(p)\rangle=\frac{1}{2}\frac{1}{p^2}\left(\eta_{\mu_1\mu_3}\eta_{\mu_2\mu_4}+\eta_{\mu_1\mu_4}\eta_{\mu_2\mu_3}\right)-\nonumber\\
&&
-\frac{D^2\a^2-2D+4}{D^2(D-2)\a^2}\frac{1}{p^2}\eta_{\mu_1\mu_2}\eta_{\mu_3\mu_4}
+\frac{2}{D-2}\frac{1}{(p^2)^2}\left(\eta_{\mu_3\mu_4}p_{\mu_1} p_{\mu_2}+\eta_{\mu_1\mu_2}p_{\mu_3} p_{\mu_4}\right)-\nonumber\\
&&-\frac{\g_1^2-4\r_1}{2\g_1^2}\frac{1}{(p^2)^2}\left(\eta_{\mu_1\mu_3}p_{\mu_2} p_{\mu_4}+\eta_{\mu_1\mu_4}p_{\mu_2} p_{\mu_3}+\eta_{\mu_2\mu_3}p_{\mu_1} p_{\mu_4}+\eta_{\mu_2\mu_4}p_{\mu_1} p_{\mu_3}\right)-\nonumber\\
&&-\frac{4(\g_1^2+2(D-2)\r_1)}{(D-2)\g_1^2}\frac{1}{(p^2)^3} p_{\mu_1} p_{\mu_2} p_{\mu_3} p_{\mu_4}
\label{Gpropaone}
\eea
}
The definition of the Feyman propagator in curves spacetime is a nontrivial issue \cite{DeWitt}\cite{Birrell}.
Maybe the cleaner appproach (because it is uniquely defined) is to consider the {\em euclidean propagator}

\par
In our case, for a plane wave  spacetime, the propagator reads
\bea &&G^E_{\m\n\r\s}\left(x,y\right)=\frac{1}{\Box}P_{\m\n\r\s}\equiv \Delta^E(x,y)\, P_{\m\n\r\s} =\nonumber\\
&&=\int_0^{\infty}d\t K \left(\tau;x,y\right)P_{\m\n\r\s}=\frac{1}{(4\pi)^{\frac{n}{2}}}\sum_{p=0}^\infty \text{tr}~ a_p(x,y)\left(\frac{ \sigma_E}{2 }\right)^{p-\frac{n}{2}+1}~\Gamma\left({n\over 2}-p-1\right)C_{\m\n\r\s}\nonumber\\\eea
where the tensor appropriate for a plane wave background, $\bg_{\m\n}$, reads
\be
\left.C_{\m\n\r\s}\equiv P_{\m\n\r\s}\right|_{\eta_{\m\n}\rightarrow \bg_{\m\n}}
\ee
and the euclidean Synge function is given by

\be \s_E(x,x_0)=\frac{1}{2}(V-V_0)(u-u_0)+\frac{1}{2}(X^a-X^a_0)C_{ab}[u,u_0](X^b-X^b_0)(u-u_0)\ee

In conclusion, in the physical dimension $n=4$ and $p=0$ (only the $a_0$ term contributes), the expression of the euclidean graviton propagator in a plane wave background reads
\be G_{\m\n\r\s}\left(x,y\right)=\frac{1}{(4\pi)^2}\frac{2}{\s_E\left(x-y\right)}C_{\m\n\r\s}\ee
What we have computed up to now is the {\em euclidean propagator.} In flat space there is a systematic way to get from it Feynman's propagator by analytic continuation, the simplest form of which is through the heat kernel approach. Assuming the same is true in curved space leads to\cite{DeWitt}\cite{Birrell}
\be
G_F(x,\xp)=-i \int_0^\infty\,ds\,e^{-i K s}
\ee
where 
\be
K\equiv |g(x)|^{-1/2}|g(x^\prime|^{-1/2}\left(\Box +m^2-i\e\right)\d(x-\xp)
\ee
and
\be
 e^{-i K s}(x,\xp)={i \Delta^{1/2}(x,\xp)\over (4\pi)^2}\,(i s)^{-n/2}\,e^{-i\left(m^2+i\e\right)s-{\s(x,\xp)\over 2 s}}F\left(x,\xp|is\right)
\ee
Finally the Schwinger-DeWitt short time expansion reads
\be
F\left(x,\xp|is\right)=a_0+a_1(is)+ a_2(is)^2+\ldots
\ee
The van Vleck-Morette determinant is defined as
\be
\Delta(x,\xp)=|g|^{-1/2}|g^\prime|^{-1/2}\,\text{det}\,\left(-\pd_\m\pd_{\m^\prime}\s(x,\xp)\right)
\ee
To give an example, when $h_{ab}$ is constant
\be
\Delta(x,x_0)=-{1\over 4}
\ee
Feynman's propagator is then
\be
G_F(x,x_0)={2 i\over (4\pi)^2}\Delta(x,x_0)^{1/2}{1\over \s(x,x_0)-i\e}={2 i\over (4\pi)^2}\Delta(x,x_0)^{1/2}\left(P{1\over \s(x,x_0)}+i\pi\d(\s)\right)
\ee
It is a well-known fact \cite{Friedlander} that when Huygens principle does not hold (like in odd-dimensional spacetimes) there is a tail term in the fundamental solution of the wave equation. This fact has been studied in detail in \cite{Ward}\cite{Adamo}. It has been claimed \cite{Harte} that in spite of the fact that in four dimensiona Huygens principle holds, this is not so for the gravitational fluctuations around a plane wave owing to a position-dependent mass-like coupling between Riemann's tensor and the fluctuations.
\par
 The fundamental solution we are talking about here refers to  what in quantum field theory is called the {\em retarded propagator} appropiate for classical computations. When calculating quantum corrections the appropiate propagator is Feynman's however.
\par
 There is a relationship \cite{Gibbons}\cite{Birrell} between the average of the advanced and retarded propagators, the Hadamard function, $G^{(1)}$ and Feynman's propagator, namely
\be
G^F_{\a\b\g\d}(x,\xp)=-{1\over 2}\left(G^{adv}_{\a\b\g\d}(x,\xp)+\,G^{ret}_{\a\b\g\d}(x,\xp)\right)-{i\over 2}\,G^{(1)}_{\a\b\g\d}(x,\xp)
\ee
where Hadamard's function is defined by
\be
G^{(1)}_{\a\b\g\d}(x,\xp)\equiv \left\langle 0\left|\left\{g_{\a\b}(x) g_{\g\d}(\xp)\right\}\right|0\right\rangle
\ee
\section{The effective action of quantum gravity in  a plane wave spacetime.}
In this section, we want to calculate the effective action of Einstein  gravity
\bea
S_{\text{\tiny{EH}}}&&\equiv -\frac{1}{2\kappa^2}\int
d^n x\sqrt{|g|}R
\eea 
It should be  expanded around a background, $\bg_{\m\n}$, which in our case it will be the plane wave
\be
g_{\m\n}=\bg_{\m\n}+\kappa h_{\m\n}
\ee
The full quadratic action is obtained after adding  the harmonic (de Donder) gauge fixing piece  \cite{AFLV}\cite{AA}
\bea
&&S_{\text{\tiny{2+gf}}}=-\frac{1}{4}\,\int\,d^nx\,\,h^{\m\n}D_{\m\n\r\s}h^{\r\s}
\eea 
where  
\be D_{\m\n\r\s}=\frac{1}{4}C_{\m\n\r\s}\Box+{1\over 2}\left(R_{\m\r\n\s}+R_{\n\r\m\s}\right)=\frac{1}{4}(\bg_{\m\r}\bg_{\n\s}+\bg_{\m\s}\bg_{\n\r}-\bg_{\m\n}\bg_{\r\s})\Box+{1\over 2}\left(R_{\m\r\n\s}+R_{\n\r\m\s}\right)\ee
This full operator that includes a massive term of sorts for some components of the graviton (the terms involving the Riemann tensor) is too complicated  for computing its inverse exactly, which would be the full propagator. Solutions
to the linear equation for the fluctuations have been provided in \cite{Harte} using the retarded Green function.
\par
What we can do is to consider the terms involving the Riemann tensor as a potential and perform a perturbative computation.
Let us recall now   that the DeWitt-Schwinger coefficients of one-loop  counterterm are given by integrals of pointwise
scalars constructed out of Riemann tensor and its covariant derivatives contracted
with the metric tensor. But in plane wave background, being  VSI, there are none of
those, so that the piece of the (euclidean) effective action not involving Riemann's tensor is 
\be
W^0_E= \log\,\det\,\Box _h-2 \log\det \Box_{gh}
\ee
where the determinant of the  Laplacian acting on scalars is given by
\be
\text{log~det}~\Box=-\int_0^{\infty}\frac{d\t}{\t}\text{tr} K \left(\tau;x,y\right)\ee
again, we directly calculate
\be
\text{log~det}~\Box=-\frac{1}{(4\pi)^{\frac{n}{2}}}\int \sqrt{|g|}~d^n x\sum_{p=0}^\infty \text{tr}~ a_p(x,y)~\lim_{\s\rightarrow 0}\left(\frac{ \sigma}{2 }\right)^{p-\frac{n}{2}}~\Gamma\left({n\over 2}-p\right)\ee
as only the $p=0$ survives 
\be
\text{log~det}~\Box=-\frac{1}{(2\pi)^{n/2}}\int \sqrt{|g|}~d^n x~\lim_{\s\rightarrow 0}
\frac{1}{ \sigma^{\frac{n}{2}} }~\Gamma\left({n\over 2}\right)~
\ee
The integrand  depends on  ${1\over \s(0)^2}$  and besides the integral is proportional to the  total (divergent) volume of the spacetime manifold. It can be interpreted as a divergent contribution to the  cosmological constant, $\l_{\infty}$.
In the presence of physical sources,
\bea
&e^{-W_E\left[T_{\m\n}\right]}=e^{-W_E[0]} \,e^{-{1\over 2}\int d(vol) \left(R_{\m\r\n\s}+R_{\n\r\m\s}\right){\d\over \d T_{\m\n}(x)}{\d\over \d T_{\r\s}(x)}}\times\nonumber\\
 &\times\, e^{,-\int d(vol)_x d(vol)_{\xp}\, T^{\m\n}(x) G^E_{\m\n\r\s}\,(x,\xp)\,T^{\r\s}(\xp)}
\eea
with
\be
W_E[0]=-{\l_\infty\over \kappa^2}\int d(vol)
\ee
and where $G^E_{\m\n\r\s}$ is the euclidean propagator  just discussed in the previous section. The first few terms in the expansion read
\bea
&e^{-W_E\left[T_{\m\n}\right]}=e^{-W_E[0]} \,\bigg\{1-{1\over 2}\int d(vol) \left(R_{\m\r\n\s}(x)+R_{\n\r\m\s}(x)\right){\d\over \d T_{\m\n}(x)}{\d\over \d T_{\r\s}(x)}+\nonumber\\
&+{1\over 4}\int d(vol)_x \left(R_{\m\r\n\s}(x)+R_{\n\r\m\s}(x)\right){\d\over \d T_{\m\n}(x)}{\d\over \d T_{\r\s}(x)}\int d(vol)_y \left(R_{\m\r\n\s}(y)+R_{\n\r\m\s}(y)\right){\d\over \d T_{\m\n}(y)}{\d\over \d T_{\r\s}(y)}+\ldots\bigg\}
\times\nonumber\\
 &\times\, e^{,-\int d(vol)_x d(vol)_{\xp}\, T^{\m\n}(x) G_{\m\n\r\s}\,(x,\xp)\,T^{\r\s}(\xp)}
\eea

The first terms in the expansion read
\bea
&e^{-W_E\left[T_{\m\n}\right]}=e^{-W_E[0]} \bigg\{- \left[R^{\m\r\n\s}(x)+R^{\n\r\m\s}(x)\right]\left(2 G^E_{\r\s\m\n}+4 G^E_{\r\s\a\b} T^{\a\b} G^E_{\m\n\g\d} T^{\g\d}+\ldots\right)\bigg\}\nonumber\\
&\times\, e^{-\int d(vol)_x d(vol)_{\xp}\, T^{\m\n}(x) G^E_{\m\n\r\s}\,(x,\xp)\,T^{\r\s}(\xp)}=e^{-W_E[0]} \Big\{1-\frac{1}{4\pi^4}\int ~d^n z \sqrt{|\bg(z)|}\bR_{\a\l\b\t}T^{\a\b}(z)T^{\l\t}(z)+\ldots\Big\}\times\,\nonumber\\
&\times e^{-\int d^n x\sqrt{|\bg(x)|} ~d^n y \sqrt{|\bg(y)|}\, \frac{1}{\s_E\left(x,y\right)}T^{\m\n}(x) C_{\m\n\r\s}\,(x,y)\,T^{\r\s}(y)}
\eea
To sum up, gravitons in this background behave almost as free particles, with a mass term of sorts which is the remembrance of the {\em tail} term in Feynman's propagator. Those are in  some sense {\em gravitational partons}. \footnote{If all we want is to compute Green functions, which are functional derivatives with respect to the sources, when the sources are equal to zero, then this can be written in a more compact form using the famous formula
\be
F\left(-i{\d\over \d x}\right) G(x)=G\left(-i{\d\over \d y}\right) \left. F(y) e^{i k(x,y)}\right|_{y=0}
\ee
The result is
\bea
&e^{-W_E\left[T_{\m\n}\right]}=e^{-W_E[0]} \times\, e^{-\int d(vol)_x d(vol)_{\xp}\, {\d\over \d h_{\m\n}}(x) G^E_{\m\n\r\s}\,(x,\xp)\,{\d \over \d h_{\r\s}}(\xp)}\nonumber\\
 &\left. \,e^{-{1\over 2}\int d(vol) \left(R_{\m\r\n\s}+R_{\n\r\m\s}\right)\,h^{\m\n}(x)\,h^{\r\s}(x)}\,e^{-\int d(vol)  h_{\m\n}T^{\m\n}}\right|_{h_{\a\b}=0}
 \eea
}

\section{The Unimodular gauge}

In the particular case of plane waves (as in all VSI spacetimes), the equations of motion of unimodular gravity are equivalent to the ones of Einstein General Relativity, because the Ricci tensor is already traceless to begin with.
As a matter of fact, in the unimodular gauge the plane wave metric, in Rosen coordinates,  reads
\be
ds^2= {2 d U d V\over\sqrt{ h[u]}}-h_{ab}[u] dX^a dX^b
\ee
where 
\be
h[u]\equiv \det \,h_{ab}[u]
\ee
and $U$ is a function of $u$ given by
\be
U(u)\equiv \,\frac{1}{2}\int^u \sqrt{ h(x)}\, dx
\ee
Next, we only need the world function for obtain the propagator and the effective action. The geodesics are determined by the Lagrangian
\be \mathcal{L}=\frac{2}{\sqrt{h[u]}}\frac{d U}{d\t}\frac{dV}{d\t}-h_{ab}[u]\frac{d X^a}{d\t}\frac{d X^b}{d\t}\ee
where $\t$ is the proper time, the momentum about $V$ yields
\bea 
&&p_V=\frac{2}{\sqrt{h[u]}}\frac{d U}{d\t}
\eea
the Euler-Lagrange equation about $V$ implies the conservation of $p_V$, then we can write
\be \frac{dU}{d\t}=\frac{\sqrt{h[u]}}{2}p_V\ee
which implies
\be \t-\t_0=\frac{2}{p_V}\mathcal{F}[U,U_0]\ee

The function $\mathcal{F}[U,U_0]$ is then given by
\be\mathcal{F}[U,U_0]\equiv\int^U_{U_0} \frac{dx}{\sqrt{h[u[x]]}}\ee
this formula yields the function $\mathcal{F}[U,U_0]$ through the implicit function theorem.

The transverse momentum can be written as
\be
p_a=-h_{ab}p_V\sqrt{h[u]}\frac{dX^b}{dU}
\ee
then
\be
X^a(\t)=X^a_0-{p_b\over  p_V}\int_{U_0}^U {h^{ab}(x)\over \sqrt{h(x)}} dx= X^a_0-{p_b\over  p_V}\mathcal{C}^{ab}[U,U_0]
\ee
where we define
\be \mathcal{C}^{ab}[U,U_0]\equiv\int_{U_0}^U {h^{ab}(x)\over \sqrt{h(x)}} dx\ee
with $\mathcal{C}_{ab}\mathcal{C}^{bc}=\d_a^c$.

Imposing now the first integral
\be
\frac{2}{\sqrt{h[u]}}\frac{d U}{d\t}\frac{dV}{d\t}-h_{ab}[u]\frac{d X^a}{d\t}\frac{d X^b}{d\t}=1
\ee
it follows that
\bea
&&V=V_0+\frac{2}{p_V^2}\mathcal{F}[U,U_0]+\frac{p_ap_b}{2p_V^2}\mathcal{C}^{ab}[U,U_0]
\eea
and world function in the unimodular gauge reads
\bea \s(x,x_0)&&=\frac{1}{2}(\t-\t_0)^2=\frac{2}{p_V^2}\mathcal{F}^2[U,U_0]=\nonumber\\
&&=(V-V_0)\mathcal{F}[U,U_0]-\frac{1}{2}(X^a-X^a_0)\mathcal{C}_{ab}[U,U_0](X^b-X^b_0)\mathcal{F}[U,U_0]\eea

It is to be concluded  that with the world function, we can obtain the propagator and the effective action in the unimodular background.
\section{Conclusions.}
In this paper we have determined the exact one loop propagator (albeit in a formal way) as well as the effective action for a graviton propagating in a plane wave. Gravitational plane waves are  roughly  similar to electromagnetic plane waves, but for the crucial point that a general gravitational wave cannot be written as some superposition of gravitational plane waves, owing to the nonlinear character of Einstein's equations. There are some exceptions to this whenever the waves are in the Kerr-Schild family in which case Ricci flatness is equivalent to Fierz-Pauli \cite{Alvarez:2024rqb}.

This work has been possible owing to the fact that plane waves belong to the family of  VSI spacetimes (all geometric scalar invariants vanish).
Then the Schwinger-DeWitt expansion is trivial, and the heat kernel corresponding to the full laplacian is exact once the world function (one half the squared geodesic distance) is known, which is just an elementary exercise. There is a quadratic coupling to the Riemann tensor, which is like a position dependent mass, and looks like the manifestation of the {\em tail terms} is this formalism.
 
 We have also pointed out that for Ricci flat spacetimes (our plane waves are such) are examples of instances where the equations of motion of unimodular gravity are the same as the general relativistic ones in the unimodular gauge.

\section{Acknowledgements}

One of us (EA) is grateful for  stimulating discussions with Luis \'Alvarez-Gaum\'e and Carmelo P. Mart\'{i}n.
 We acknowledge partial financial support by the Spanish MINECO through the Centro de excelencia Severo Ochoa Program  under Grant CEX2020-001007-S  funded by MCIN/AEI/10.13039/501100011033.
 We also acknowledge partial financial support by the Spanish Research Agency (Agencia Estatal de Investigaci\'on) through the grant PID2022-137127NB-I00 funded by MCIN/AEI/10.13039/501100011033/ FEDER, UE.
All authors acknowledge the European Union's Horizon 2020 research and innovation programme under the Marie Sklodowska-Curie grant agreement No 860881-HIDDeN and also byGrant PID2019-108892RB-I00 funded by MCIN/AEI/ 10.13039/501100011033 and by ``ERDF A way of making Europe''.

\newpage
\appendix
\section{VSI spacetimes}\label{A}
Let us summarize here   some definitions purporting on  spacetimes with parallel rays in General Relativity \cite{Englert}, \cite{Roche:2022bcz} and \cite{Blau}.

{\it Parallel Wave}

A parallel wave is a Lorentzian manifold $(M,g)$ which admits a global, covariantly constant, null vector field $Z$.
\be \nabla_\n Z_\m=0\ee
the "rays" of such a wave are automatically (null) geodesics since $Z$ is covariantly constant.

This means \cite{Gibbons}, \cite{Galaev} that the holonomy of the manifold is the little group of the null manifold generated by $Z$, that is, the similitude group $\mathfrak{sim}(2)=\left(\mathbb{R}\oplus \mathfrak{so}(n)\right)\ltimes \mathbb{R}^2$, which is, in some sense, the maximal subalgebra  of the Lorentz $SO(1,3)$ one. It is amusing to remark that this is the subalgebra that yields Cohen and Glashow's {\em very special relativity} \cite{CohenG}, \cite{Alvarez:2008uy}, generated by
\be
\bigg\{T_1\equiv K_1+J_2,\quad T_2\equiv K_2-J_1,\quad J_3,\quad K_3\bigg\}
\ee
where $K_i (1=1,2,3)$ generate boosts and $J_i (1=1,2,3)$ generate ordinary rotations.

This definition immediatly inplies that $Z$ is a Killing. The Ricci identity implies that $Z$ is an eingenvector of Riemann tensor with eigenvalue zero
\be
R_{\m\n\r\s} Z^\s=0
\ee
which is Bel's criteria \cite{Bel} for the spacetime to be type N.

{\it Wavefront of a Parallel Wave}

If a parallel wave is definied by a covariantly constant null vector field $Z$, then the wavefront of such a wave is defined as the ortogonal space  
\be Z_{\perp}\slash Z\ee
where $Z_{\perp}\equiv\left\{x\in TM\mid g(x,Z)=0\right\}$

{\it Plane-fronted Wave with Parallel Rays (pp-Wave)}

A pp-wave is Lorentzian manifold $(M,g)$ which admits a global, covariantly constant, null vector field $Z$, in which the curvature tensor satisfies 
\be R\mid_{Z_{\perp}\land Z_{\perp}} =0\ee
this means that $Z$ is not only a Killing vector field, but that  also it is a gradient. If it does not vanish anywhere, we can define a null coordinate such that
\be
Z\equiv{\pd \over \pd v}
\ee
To be specific, the most general metric admitting a covariantly constant null vector can be written as
\be
\text{d}s^2=\text{d}u\left(\text{d}v-F[u,x^c]\text{d}u-G_a[u,x^c]\text{d}x^a\right)-g_{ab}\text{d}x^a\text{d}x^b
\ee
where the light cone coordinates
\bea
&&u\equiv x^{-}= t-z\nonumber\\
&&v\equiv x^{+}=t+z
\eea
and the transverse coordinates, are given by ($a,b=1,2$)
\be
x_T\equiv (x^c)\equiv (x,y)\quad 
\ee
The {\it vanishing curvature invariants}  metric \cite{Coley:2008th} reduces to
\be
\text{d}s^2=\text{d}u\left(\text{d}v-F[u,x^c]\text{d}u-G_a[u,x^c]\text{d}x^a\right)-\d_{ab}\text{d}x^a\text{d}x^b
\ee
in particular the standard {\it pp-wave} metric reads
\be
\text{d}s^2=\text{d}u\text{d}v-F[u,x^c]\text{d}u^2-\d_{ab}\text{d}x^a\text{d}x^b
\ee

Finally, a {\it plane wave} is a classical pp-wave for which the characteristic function $F$ is quadratic in $x^c$, $F[u,x^c]=H_{ab}x^ax^b$
\be
\text{d}s^2=\text{d}u\text{d}v-H_{ab}[u]x^ax^b\text{d}u^2-\d_{ab}\text{d}x^a\text{d}x^b
\ee
in harmonic (often called Brinkmann) coordinates. 

Note that the wave fronts 
\be
u=C
\ee
are flat, then planar. The other part of the name (parallel rays) refers to the existence of a parallel null vector. 
Shifts of the coordinate $v$ 
\be
\d v=\Lambda(u,x_T)
\ee
belong to the residual gauge symmetry. Then
\bea
&&\d F={1\over 2}\pd_a\Lambda\nonumber\\
&&\d G=\pd_a \Lambda
\eea

  
\end{document}